# Performance Enhancement of Nano-Scale VO$_2$ Modulators using Hybrid Plasmonics

Herman M. K. Wong and Amr S. Helmy, *Senior Member, IEEE*

*Abstract*— Vanadium dioxide (VO$_2$) is a phase change material (PCM) that exhibits a large change in complex refractive index on the order of unity upon switching from its dielectric to its metallic phase. Although this property is key for the design of ultra-compact optical modulators of only a few-microns in footprint, the high absorption of VO$_2$ leads to appreciable insertion loss (IL) that limits the modulator performance. In this work, through theory and numerical modeling, we report on a new paradigm, which demonstrates how the use of a hybrid plasmonic waveguide to construct a VO$_2$ based modulator can improve the performance by minimizing its IL while achieving high extinction ratio (ER) in comparison to a purely dielectric waveguide. The hybrid plasmonic waveguide that contains an additional metal layer with even higher loss than VO$_2$ enables unique approaches to engineer the electric field (E-field) intensity distribution within the cross-section of the modulator. The resulting Figure-of-Merit FoM = ER/IL is much higher than what is possible by simply incorporating VO$_2$ into a silicon wire waveguide. A practical modulator design using this new approach, which also includes input and output couplers yields ER = 3.8 dB/μm and IL = 1.4 dB/μm (FoM = 2.7), with a 3-dB optical bandwidth >500 nm, in a device length = 2 μm, and cross-sectional dimensions = 200 nm × 450 nm. To our knowledge, this is one of the smallest modulator designs proposed to-date that also exhibits amongst the highest ER, FoM, and optical bandwidth, in comparison to existing designs. In addition to VO$_2$, we investigate two other PCMs incorporated within the waveguide structure. The improvements obtained for VO$_2$ modulators do not extend to other PCMs.

## I. INTRODUCTION

IN modern long-haul and data communications, optical modulators are an indispensable component, due to the numerous advantages of optical communications such as ultra-high data capacity, low loss transmission, and the signal is not affected by external electromagnetic interference. Emerging optical communications applications are moving towards the dense integration of optoelectronic devices to increase system performance while reducing both cost and energy consumption [1-3]. As such, there is an urgent need to further reduce the footprint of individual photonic devices, such as the modulator.

Phase change materials (PCMs) have emerged as promising candidates for implementing photonic devices, due to the ability to substantially transform their physical properties using a minute change in external stimulus. The optical switching associated with the metal-insulator transition (MIT) of vanadium dioxide (VO$_2$), for example, has been proposed for numerous applications. VO$_2$ has been integrated with micro- and nanophotonic structures to achieve useful devices, such as waveguide modulators for optical communications [4], tunable plasmonics based optical antennas to realize spatial light modulators or active displays [5], ultrathin perfect absorbers [5], and optical gating for light triggered thyristors in power electronics [6]. Another class of PCMs is the chalcogenides, such as Ge$_2$Sb$_2$Te$_5$ (GST), which are capable of switching between amorphous and crystalline states on a picosecond timescale. The property of GST to retain a given material state for an extended period of time, which can be controlled even at the nanoscale, is of great utility for data storage devices such as optical discs and random access memories [7]. More recently, on-chip photonic memory elements utilizing GST have been experimentally demonstrated [8].

PCMs such as VO$_2$ are advantageous for optical waveguide modulators because of the large change in complex refractive index upon the material phase change between its two different states. This change is orders of magnitude larger than what is achievable using typical electro-optic [9], electro-absorption [10], and free carrier dispersion [11] effects. For example, the common technique of injecting free carriers in semiconductors induces refractive index changes on the order of $10^{-2}$ [12], but the change in refractive index in PCMs is on the order of unity. This substantial change in PCMs can lead to modulators with device lengths significantly shorter than what is attainable using more conventional active materials.

However, the common drawback of using PCMs for optical modulation is their high material absorption. For a modulator whose main guiding layer is a PCM, the insertion loss (IL), which is the device loss in the ON-state, would still be quite significant (i.e., >10 dB/μm at λ = 1.55 μm). To reduce the loss of a waveguide implemented with a PCM, the PCM must only be a relatively thin layer within a host waveguide. However, this in turn compromises the extinction ratio (ER) of a modulator implemented using this structure. ER is defined as the difference in loss between the ON- and OFF-state of the modulator. This presents a tradeoff between the insertion loss and the attainable extinction ratio.

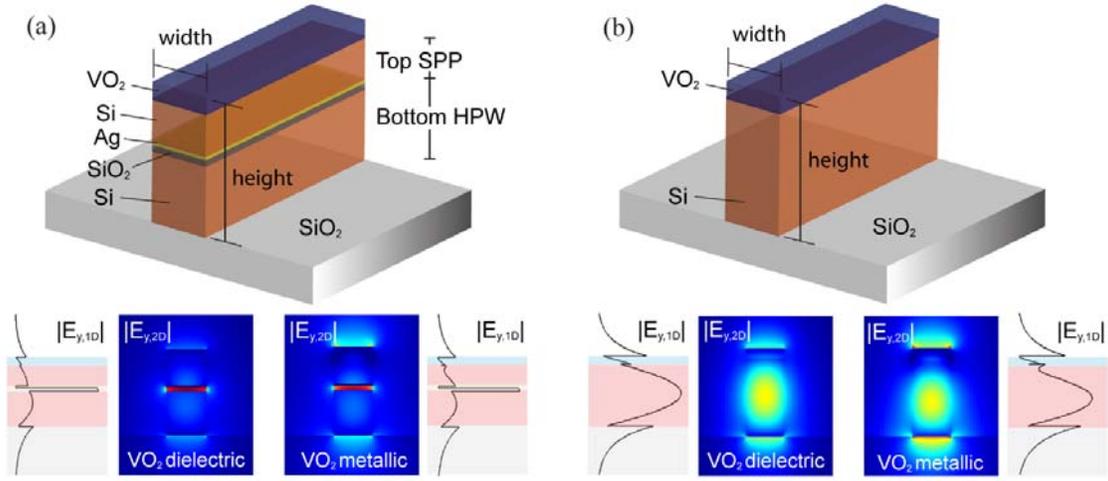

Fig. 1. 3D Schematic of the (a) HPSPP/VO$_2$ and (b) Si wire/VO$_2$ waveguide showing the different material layers, and the E$_y$ field amplitude distribution from both 1D and 2D mode simulations for the cases of when VO$_2$ is dielectric and metallic.

In this paper, it will be shown through theory and numerical modeling that by utilizing a unique hybrid plasmonic waveguide structure, the performance of a modulator that uses VO$_2$ can be significantly improved when compared to a device that is based on a total-internal-reflection (TIR) waveguide with VO$_2$ incorporated. The IL of the hybrid plasmonic modulator that contains an additional lossy metal layer can be even lower than that of the dielectric TIR modulator; this is somewhat counterintuitive but is attained through electric-field (**E**-field) engineering. Firstly, the design and optimization of a VO$_2$ based hybrid plasmonic modulator with high performance over a broad wavelength range of >500 nm, in a device length of only 2 μm is shown. Our modulator design exhibits the highest optical bandwidth in comparison to previously demonstrated devices based on a range of platforms and active materials [4,13-22], in a device footprint that is amongst the smallest, and both the ER and FoM are also higher than with most other devices. After the design of our hybrid plasmonic modulator is shown, the **E**-field engineering formalism and design equations of the VO$_2$ hybrid plasmonic modulator are elucidated. Furthermore, PCMs other than VO$_2$ are investigated for use as the switching material within the hybrid plasmonic modulator, and the performance is compared to VO$_2$ based modulators.

## II. DESIGN OF VO$_2$ HYBRID PLASMONIC MODULATOR

The modulator design is based on a unique hybrid plasmonic waveguide structure (Fig. 1a) that consists of 4 material layers on top of SiO$_2$ substrate excluding the VO$_2$. In our waveguide design as shown in Fig. 1a, the high index layer is Si, the low index layer is SiO$_2$, and the metal layer is chosen to be Ag as determined by limiting factors in the planned fabrication process. The thickness of the bottom Si layer is 220 nm as it is the standard Si thickness of a Silicon-on-Insulator (SOI) wafer, and the thicknesses of the SiO$_2$ and Ag are 20 nm and 10 nm, respectively. The fundamental transverse magnetic (TM) mode in this structure is a symmetric mode that is the coupled mode between the surface plasmon polariton (SPP) mode at the single-interface between the top high index layer and metal layer, and the conventional hybrid plasmonic waveguide (HPW) mode formed by the lower high index/low index/metal layers [23]. Here, we refer to our waveguide structure as the hybrid plasmonic surface plasmon polariton (HPSPP) waveguide, which was first proposed and demonstrated in Ref. 24. The main advantage of the HPSPP is that by tuning the **E**-field components at either side of the metal layer, the propagation loss of the symmetric mode can be optimized to be of a very low value on the order of $10^{-2}$ dB/μm. The remarkable property of this waveguide is that the lower losses are attainable while simultaneously maintaining the mode area to be similar to that of the conventional HPW. The HPSPP is a modification of the asymmetric hybrid plasmonic waveguide (AHPW) from Ref. 25, where there are 5 material layers such that the fundamental TM mode is a coupled mode between the modes of the top and bottom HPW.

By adding a thin layer of VO$_2$ on top of the HPSPP structure, a modulator is formed, where the ON-state is achieved when the VO$_2$ is a dielectric (high transmission), and upon switching VO$_2$ into its metallic state, the OFF-state is induced (low transmission). Modulation is activated by the large change in complex refractive index of VO$_2$ when it transitions from dielectric to metallic phase; namely from $n_d = 2.861 + 0.267i$ in the dielectric state to $n_m = 1.664 + 3.291i$ in the metallic state at the operating wavelength λ = 1.55 μm. These values are obtained from ellipsometry measurements of polycrystalline films deposited by pulsed laser deposition, which we utilize in this work; we also employ the measured wavelength-dependent complex refractive indices for broadband mode solving and finite-difference time-domain (FDTD) simulations. A comprehensive set of values of refractive indices of VO$_2$ in its dielectric and metallic states can be found in Ref. 26. The rationale behind implementing the modulator based on the HPSPP waveguide structure is that it enables much higher losses when VO$_2$ is in its metallic state compared to the case based on a simple total-internal-reflection (TIR) waveguide (Fig. 1b), while simultaneously provides comparable losses when the VO$_2$ is in its dielectric state. In essence, this architecture enables the ON-state loss, which is also termed insertion loss (IL) to be maintained at relatively low values while significantly increasing the difference between the OFF- and ON-state losses, or extinction ratio (ER). The deposition of VO$_2$ on a variety of different substrates including Si has been demonstrated in previous work [4, 6, 16, 17], and thus the HPSPP/VO$_2$ modulator structure we present here is amenable to fabrication with current technologies.

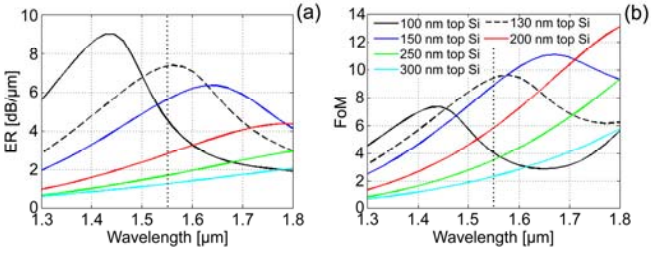

Fig. 2. (a) Extinction ratio and (b) Figure-of-Merit of HPSPP/VO$_2$ modulator as a function of wavelength for different top Si (high index) layer thicknesses when VO$_2$ thickness is 50 nm (2D mode solving). Vertical dotted line indicates operating wavelength.

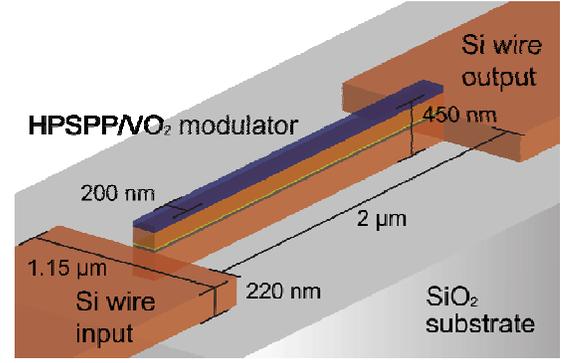

Fig. 3. Schematic of HPSPP/VO$_2$ modulator of 2-µm length including input and output Si wire waveguide couplers. The optimized critical dimensions of the device are shown.

The HPSPP allows for the guiding of a TM symmetric mode in which the EM field intensity within the lossy VO$_2$ layer is much lower than in the case when a TIR waveguide is used (e.g., VO$_2$ layer on top of a Si wire waveguide). This is because the hybrid plasmonic mode leads to concentration of EM field within the low index layer (Fig. 1a), and thus the field is pulled away from the VO$_2$ layer. Although the metal layer causes additional loss, the thicknesses of the HPSPP waveguide layers can be tuned to minimize the EM field inside the metal, thus minimizing loss [24]. When VO$_2$ is switched to become metallic, its material loss increases dramatically. In addition, the EM field intensity also increases within the layer (Fig. 1) due to the change in complex refractive index. As such, the combination of these two effects leads to a significant increase in the overall waveguide loss.

## A. Device Optimization and Performance

The goal of the waveguide design process is to maximize the extinction ratio (ER) defined as the difference in loss between the ON and OFF-states, while also keeping the loss of the ON-state or insertion loss (IL) very low. An effective Figure-of-Merit (FoM) for a modulator to determine its performance is FoM = ER/IL. The first step of the waveguide design process is to determine ER, IL, and FoM as a function of the top Si layer thickness, while keeping the thicknesses of all other HPSPP waveguide layers constant (50 nm VO$_2$). This is achieved by 2D mode solving to obtain the modulator performance based on modal losses, for the purpose of device optimization with reasonable computational times. Next, the wavelength-dependent device performance for several different top Si layer thicknesses is determined also by 2D mode solving, to observe trends in both the optical bandwidth and optimal wavelength for peak performance. Lastly, 3D FDTD simulations are utilized to study the HPSPP/VO$_2$ modulator structure optimized with 2D mode solving to obtain more detailed performance characteristics. In general, 2D mode solving is utilized for optimization in which different waveguide layer thicknesses are tuned, while 3D simulation is for analyzing the broadband performance of realistic structures to be implemented experimentally.

The results of optimization by 2D mode solving using Lumerical MODE [28] at λ = 1.55 µm shows that the best performance of our modulator is achieved when the top Si thickness is 130 nm and the VO$_2$ layer is 50 nm thick, such that both ER = 7.4 dB/µm and FoM = 9.4 are maximized. Note that the Palik model [27] is utilized for the wavelength-dependent complex refractive index of Ag in the simulations. To study the optical bandwidth of the HPSPP/VO$_2$ modulator as a function of top Si thickness (with 50 nm VO$_2$ thickness), 2D mode solving is performed as a function of wavelength for different top Si thicknesses. In Fig. 2a, it can be seen that with the optimized top Si thickness of 130 nm for operation wavelength of 1.55 µm (when both ER and FoM are at their peaks), the 3-dB bandwidth is >300 nm. As the top Si thickness is increased, the optical bandwidth steadily increases as well, to >500 nm for top Si thicknesses of over 150 nm. Also, by increasing the top Si thickness, the peak for each of the ER and FoM shifts to a higher wavelength, the peak ER decreases (Fig. 2a), and the peak FoM increases (Fig. 2b).

The broadband characteristics of the HPSPP/VO$_2$ modulator are investigated in more details using full-wave 3D FDTD simulation [28]. Given the sizes of these modulators, input and output couplers from and into passive Si wire waveguides must be designed [24] (Fig. 3). The coupling efficiency is maximized by matching the effective indices ($n_{eff}$) of the Si wire and HPSPP/VO$_2$ waveguide modes; the coupling loss originates from the mismatch in modal overlap between the Si wire TM mode and the TM symmetric mode of the HPSPP/VO$_2$. When the input and output couplers are included in a 3D FDTD simulation, ER = 3.8 dB/µm and FoM = 2.7 are obtained at the operating wavelength λ = 1.55 µm when the top Si thickness of the HPSPP/VO$_2$ is optimized to be 150 nm (with 50 nm VO$_2$ thickness), which deviates from that obtained using 2D mode solving. The optimized Si wire coupler waveguide has a width of 1.15 μm and height of 220 nm to match the SOI Si layer thickness (Fig. 3). It is found that the coupling efficiency from the optimized Si wire waveguide into the HPSPP/VO$_2$ modulator reaches ~80% when VO$_2$ is in its dielectric phase, which is desirable as a high performance modulator requires high power transmission at its ON-state. On the other hand, it is not useful to achieve high coupling efficiency when VO$_2$ is in its metallic phase, as it is actually better for the OFF-state transmission to be low. A schematic of the HPSPP/VO$_2$ modulator with input and output Si wire waveguide couplers is shown in Fig. 3. Plots of the IL, ER, FoM, and reflection as a function of wavelength are shown in Fig. 4, based on a 2-μm length HPSPP/VO$_2$ modulator coupled into and out of by Si wire waveguides (Fig. 3). The 3-dB optical bandwidth of the modulator is >500 nm (Fig. 4b). Note that a similar bandwidth using a HPSPP structure was first achieved in Ref. 24. The IL varies from ~1.1 to 1.5 dB/μm. The reflection at the ON-state of the modulator shows a clear set of fringes due to multiple Fabry-Perot reflections within the modulator section. At the operating wavelength of 1.55 μm, the reflection is quite low at ~ -23 dB, which is advantageous when the modulator is placed close to a laser source. When VO$_2$ is switched to metallic for the OFF-state, the reflection fringes are less prominent because of the increased loss of the HPSPP/VO$_2$ waveguide; the average reflection is ~ -16 dB. The results of 3D FDTD simulations show a significantly reduced ER and FoM, and higher IL compared to that obtained from 2D mode solving, with the optimized top Si thickness also different. This is attributed to the fact that in the realistic structure including input and output couplers, light is also coupled into waveguide modes other than the TM symmetric mode of the HPSPP/VO$_2$ modulator, such as the TM antisymmetric mode and other leaky modes.

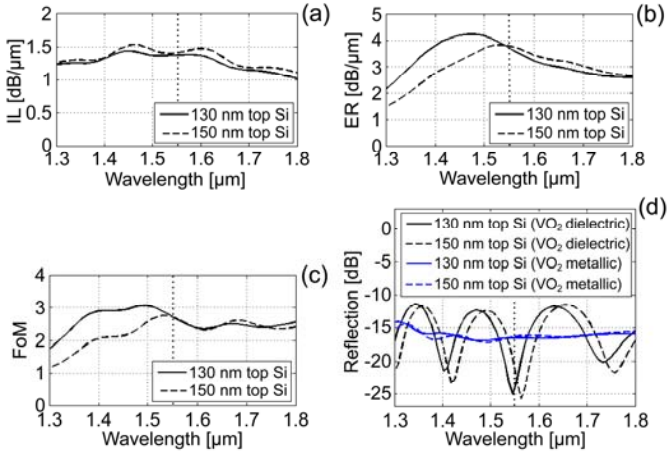

Fig. 4. (a) Insertion loss when VO$_2$ is dielectric, (b) extinction ratio, (c) Figure-of-Merit, and (d) reflection of HPSPP/VO$_2$ modulator as a function of wavelength for top Si (high index) layer thicknesses of 130 nm and 150 nm, and VO$_2$ thickness of 50 nm (3D FDTD). Vertical dotted line indicates operating wavelength.

Fig. 4 also shows the performance of the modulator when the top Si thickness is 130 nm (the optimized thickness based on mode solving). It can be seen that the peak of ER and FoM is at a lower wavelength compared to when the top Si thickness is 150 nm. With top Si thickness reduced to 130 nm, the IL is lower over the entire wavelength range. However, the ER = 3.7 dB/μm is slight lower at λ = 1.55 μm, and the FoM remains the same at ~2.7. The ER peaks at λ = 1.478 μm with a value of ~4.2 dB/μm, and the 3-dB bandwidth is ~380 nm. The reflection spectrum has shifted in terms of wavelength for both cases when VO$_2$ is dielectric and metallic. In order to shift the peak of both the ER and FoM to the desired operating wavelength λ = 1.55 μm, the top Si thickness must be increased to 160 nm. However, in this case, both the ER ≈ 3.5 dB/μm and FoM ≈ 2.4 are appreciably reduced compared to the values with other studied thicknesses of top Si. In summary, setting the top Si thickness to be 150 nm produces the best performing modulator at the operating wavelength of λ = 1.55 μm, for the case when VO$_2$ thickness is 50 nm.

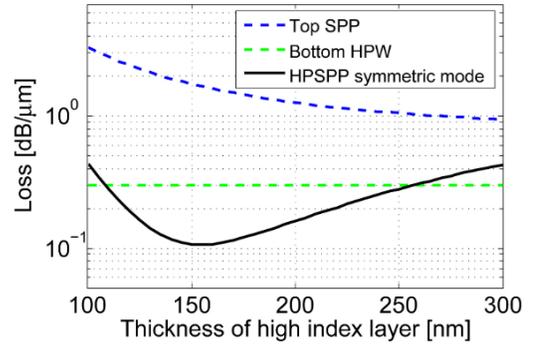

Fig. 5. Modal loss of HPSPP/VO$_2$ waveguide as a function of top Si (high index) layer thickness compared to modal losses of top SPP and bottom HPW at λ = 1.55 μm (2D mode solving).

It is expected that the modulation speed of the proposed HPSPP/VO$_2$ modulator in this work can achieve 1 GHz for electrical switching [29], and up to 100's of GHz for ultrafast optical excitation [17], based on already experimentally demonstrated VO$_2$ based modulators. The VO$_2$ phase transition from its dielectric to rutile metallic state occurs in <100 fs using ultrafast optical excitation, and the relaxation from metallic back to dielectric phase is on the order of a few picoseconds with optical fluences below a critical value of ~5 mJ/cm$^2$ [30,31]. For both electrical and optical modulation, the limiting factor in speed is thus the relaxation time it takes for VO$_2$ to transition from its rutile metallic back to dielectric phase after excitation, which can possibly be improved by incorporating an electrical bias to rapidly sweep out excess carriers; this is a proven technique in silicon based carrier depletion type optical modulators [32]. The switching energy for our proposed modulator can also be estimated. Based on the measured threshold pump fluence of ~0.25 mJ/cm$^2$ for switching nanoscale VO$_2$ from its dielectric to metallic phase pumped at a wavelength of 1550 nm [33], the footprint of the VO$_2$ strip in our design of 2 μm × 200 nm, and that there is a 1/4 likelihood of ON- to OFF-state transition in a random signal, the minimum energy required for ultrafast all-optical switching of our HPSPP/VO$_2$ modulator is ~250 fJ/bit (~1/4 ×

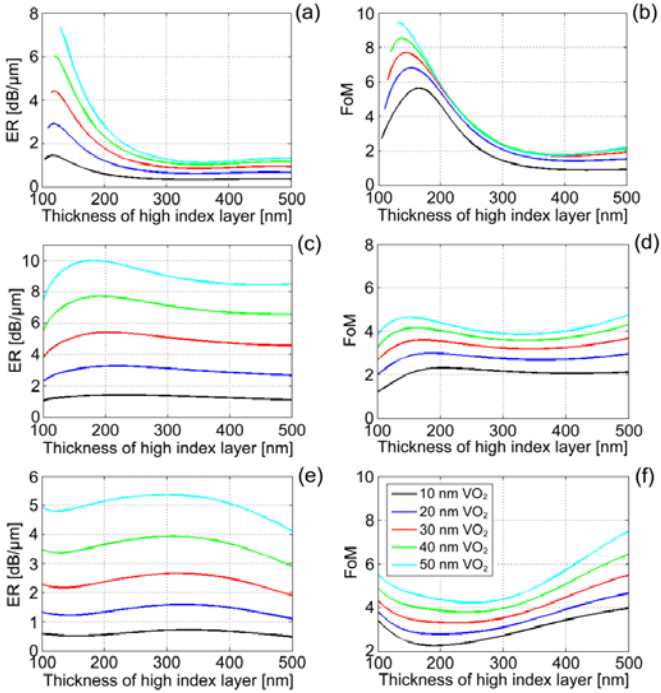

Fig. 6. Extinction ratio and Figure-of-Merit of HPSPP/VO$_2$ waveguide as a function of top Si (high index) layer thickness for different VO$_2$ thicknesses at $\lambda$ = 1.55 μm; (a, b) VO$_2$ utilized as the top layer, (c, d) VO$_2$ utilized as the low index gap layer, and (e, f) VO$_2$ utilized as the metal layer (2D mode solving).

1 pJ). This switching energy is only ~1.3 times that of the VO$_2$ on silicon micro-ring resonator based modulator demonstrated in Ref. 17, which is ~190 fJ/bit. However, our design exhibits an ultra-high optical bandwidth of >500 nm, compared to the bandwidth of the micro-ring resonator modulator of only ~1 nm based on the full-width half-maximum (FWHM) of the resonant transmission dips [17]. The broadband nature of our device means that temperature stabilization is not required during operation, and it would also be useful within systems that require multiple wavelength channels (e.g., in wavelength division multiplexing). Furthermore, while the micro-ring modulator in Ref. 17 with a diameter of 3 μm can achieve an ER ≈ 10 dB, a 3-μm length HPSPP/VO$_2$ modulator proposed in this work can achieve ER ≈ 11.4 dB.

### B. Merits of Modulator Design

Our particular HPSPP/VO$_2$ modulator design outperforms structures using the same materials and that are similar in cross-sectional dimensions. It was already shown that in Ref. 25 that the loss of the 5-layer AHPW can be engineered to be much lower than that of the conventional HPW. Similarly, our HPSPP design can also be tuned to have a lower loss than the HPW by varying the thickness of the top Si (high index) layer, as shown in Fig. 5 for the HPSPP with dielectric VO$_2$ on top [24]. Incorporating VO$_2$ into the HPSPP structure can effectively make it into a modulator, but the question becomes: where should it be inserted within the waveguide layers? The 3 options are: (1) VO$_2$ as the top layer, (2) VO$_2$ as the low index gap layer, and (3) VO$_2$ as the metal layer. By investigating the loss of the HPSPP/VO$_2$ waveguide as a function of both the top Si and VO$_2$ thicknesses using 2D mode solver at the operating wavelength of $\lambda$ = 1.55 μm, the maximum ER = 7.4 dB/μm and FoM = 9.4 are achieved when the top Si thickness is 130 nm and the VO$_2$ thickness is 50 nm, as shown in Fig. 6a and 6b for the case when VO$_2$ is placed on top of the HPSPP. However, when VO$_2$ is utilized as the low index gap layer (replacing SiO$_2$), although the maximum ER = 10.0 dB/μm when top Si thickness = 180 nm and VO$_2$ thickness = 50 nm (Fig. 6c), the maximum FoM is much lower at 4.6 when top Si thickness = 150 nm and VO$_2$ thickness = 50 nm (Fig. 6d). When VO$_2$ is used as the metal layer, the maximum ER = 5.4 dB/μm when top Si thickness = 300 nm and VO$_2$ thickness = 50 nm (Fig. 6e); for these waveguide dimensions, the FoM = 4.4 (Fig. 6f). Since FoM is the more important parameter that considers both ER and IL, the best choice for a modulator design would be for VO$_2$ to be the top layer of the HPSPP/VO$_2$ structure, which achieves a FoM that is over two times higher than by the other structures.

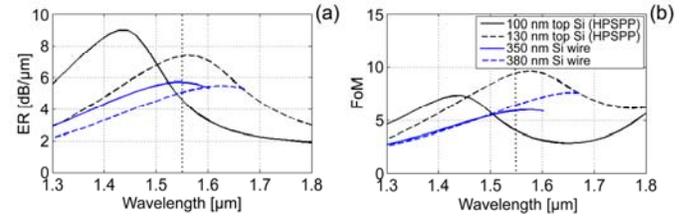

Fig. 7. (a) Extinction ratio and (b) Figure-of-Merit of HPSPP/VO$_2$ and Si wire/VO$_2$ modulators as a function of wavelength for different top Si layer or total Si wire thicknesses when VO$_2$ thickness is 50 nm (2D mode solving). Vertical dotted line indicates operating wavelength.

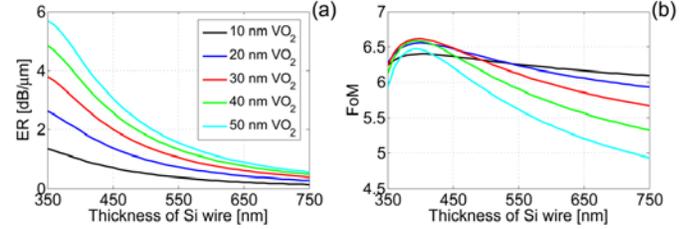

Fig. 8. (a) Extinction ratio and (b) Figure-of-Merit of Si wire/VO$_2$ modulator as a function of Si wire thickness for different VO$_2$ thicknesses at $\lambda$ = 1.55 μm (2D mode solving).

It is also important to compare our modulator design to the case when VO$_2$ directly sits on top of a Si wire waveguide (Fig. 1b), which would justify the use of a much more complex multilayered structure. In Fig. 7, comparisons between ER and FoM of the HPSPP/VO$_2$ and the Si wire/VO$_2$ modulators are shown. In the comparison, both structures have the exact same dimensions, such that they both have the same width of 200 nm, and the total height is also the same. For example, when the top Si thickness = 130 nm for the HPSPP/VO$_2$, the corresponding height of the Si wire in the Si wire/VO$_2$ is 380 nm, making the total height of each of the structures 430 nm when the VO$_2$ thickness is also considered. Fig. 7 shows that when the HPSPP/VO$_2$ modulator is optimized to operate at $\lambda$ = 1.55 μm (i.e., using 130 nm top Si thickness), both ER and FoM are significantly higher for the HPSPP compared to the Si wire based modulator over all wavelengths from 1.3 to 1.8 μm. As an example, when the top Si thickness = 130 nm for the HPSPP/VO$_2$, ER ≈ 7.4 dB/μm; however, the Si wire/VO$_2$ with the same total device height exhibits ER ≈ 5.1 dB/μm. Similarly, FoM for the HPSPP/VO$_2$ modulator is ~9.4, but FoM is only ~6.4 for the Si wire/VO$_2$ modulator. From Fig. 7, it can also be observed that the Si wire/VO$_2$ waveguide TM mode becomes cut off at wavelengths above 1.6 μm and 1.67 μm when the Si thickness is 350 nm and 380 nm, respectively; whereas the HPSPP/VO$_2$ waveguide TM symmetric mode is still guided for wavelengths above 1.8 μm when the overall waveguide dimensions are the same. This implies that the bandwidth of the Si wire/VO$_2$ modulator is in general limited to <100 nm above the operating wavelength of 1.55 μm, while for the HPSPP/VO$_2$ modulator, the bandwidth extends to much higher wavelengths.

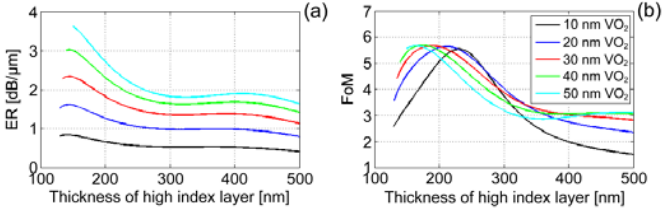

Fig. 9. (a) Extinction ratio and (b) Figure-of-Merit of AHPW/VO$_2$ modulator as a function of top Si (high index) layer thickness for different VO$_2$ thicknesses at $\lambda = 1.55$ μm (2D mode solving).

To generalize the argument that the HPSPP/VO$_2$ is better than Si wire/VO$_2$ in terms of modulator performance, the FoM of the Si wire/VO$_2$ modulator must be investigated for a range of Si wire thicknesses to determine the maximum achievable value, and the corresponding ER is also noted. Fig. 8 shows the ER and FoM of the Si wire/VO$_2$ modulator for different Si wire and VO$_2$ thicknesses. It is observed that the maximum FoM of ~6.6 is achieved when the Si wire thickness is 400 nm and VO$_2$ thickness is 30 nm, and the corresponding ER is ~2.9 dB/μm. Although this maximum FoM is slightly higher than that achieved when the Si wire and VO$_2$ layer are 380 nm and 50 nm thick, respectively, the ER is simultaneously much lower. However, regardless of any Si wire thicknesses utilized, both the FoM and ER cannot reach close to the maximum attained by the HPSPP/VO$_2$ modulator with VO$_2$ as the top layer.

Compared to the 5-layer AHPW introduced in Ref. 25, the performance of the 4-layer HPSPP is significantly better when implemented as a modulator by incorporating VO$_2$. Fig. 9 shows the ER and FoM of a 5-layer AHPW/VO$_2$ modulator consisting of the same layers as our 4-layer HPSPP/VO$_2$ modulator, but with the addition of a 20 nm SiO$_2$ low index layer on top of the Ag layer. The maximum FoM achieved when VO$_2$ is 50 nm thick is 5.7 at the top Si layer thickness of 165 nm, and the corresponding ER = 3.4 dB/μm, which are much lower than those of the HPSPP/VO$_2$ modulator as shown previously. Table 1 compares the performance of different modulator designs utilizing VO$_2$ as discussed in this section.

Finally, our HPSPP/VO$_2$ design attains modulator performance that is better than most existing devices in literature which are aimed at amplitude modulation in the telecom wavelengths, across a range of device types and active materials utilized, which is shown in Table 2. It can be observed that our modulator design is favorable in terms of all of the considered performance metrics compared to most other devices reported. Note that the majority of the listed devices from literature have been experimentally investigated, which means that a direct comparison with our theoretical study here is not completely fair. Nonetheless, the results of our proposed HPSPP/VO$_2$ modulator specifically from 3D FDTD simulations give a good indication of the theoretical best performance that can be achieved by our modulator design (Table 2). In comparison to the Si wire/ITO modulator in Ref. 15 that was theoretically investigated using 3D FDTD, our device exhibits a lower ER; however, the FoM is comparable, and the optical bandwidth of the HPSPP/VO$_2$ modulator is almost 5 times higher (Fig. 4). Compared to the experimentally demonstrated ITO based plasmonic slot waveguide modulator in Ref. 20, our design has higher IL and lower FoM, but the ER is conversely much higher (Fig. 4). For the hybrid plasmonic ITO modulator in Ref. 21 whose performance metrics are obtained from 2D mode solving, it can be seen that while our device exhibits higher IL and lower FoM, it again achieves a much higher ER (Fig. 4). Note that ER is ultimately the more important attribute for modulators once IL reaches lower than a certain level (i.e., one that is acceptable in terms of contribution to system loss). For example, in typical semiconductor integrated optical modulators, a 3-dB IL is considered to be a very good value [34], and even 5 to 6-dB IL [35, 36] is considered quite low. Our proposed HPSPP/VO$_2$ modulator with 2-μm length achieves IL ≈ 2.8 dB without considering input and output coupling, and IL ≈ 4.8 dB if coupling losses are included, which is quite acceptable.

### III. Field Engineering and Design Equations

As shown in Table 1, the use of the HPSPP waveguide structure to replace the Si wire for the VO$_2$ modulator (with VO$_2$ as the top layer) can significantly improve both the ER and FoM of the device, with IL also slightly lowered. In this section, we describe the formalism behind engineering of the EM field distribution using the HPSPP waveguide and how it is used to increase the overall waveguide loss when VO$_2$ is metallic, while maintaining low loss when VO$_2$ is dielectric.

Table 1. Comparison of insertion loss, extinction ratio, and Figure-of-Merit of different VO$_2$ based modulators investigated in this paper (2D mode solving).

| Modulator device | Top Si [nm] | IL [dB/μm] | ER [dB/μm] | FoM |
|---|---|---|---|---|
| HPSPP/VO$_2$ (VO$_2$ as top layer) | 130 | 0.78 | 7.36 | 9.41 |
| HPSPP/VO$_2$ (VO$_2$ as low index layer) | 150 | 2.11 | 9.78 | 4.63 |
| HPSPP/VO$_2$ (VO$_2$ as metal layer) | 200 | 1.23 | 5.37 | 4.38 |
| Si wire/VO$_2$ | 380 | 0.80 | 5.12 | 6.43 |
| AHPW/VO$_2$ | 165 | 0.60 | 3.39 | 5.68 |

Table 2. Comparison of amplitude modulation performance at telecom wavelengths for different device types and active materials utilized that have been reported. **Not specified

| Device type | Active material | Experiment / theory | Switching mechanism | Operation wavelength [μm] | ER [dB/μm] | IL [dB/μm] | FoM | Device length [μm] | Optical bandwidth [nm] | Energy/bit [fJ] | Speed [GHz] |
|---|---|---|---|---|---|---|---|---|---|---|---|
| Si nanowire [13] | InGaAlAs MQW | Experiment | Electrical | 1.5 | 0.098 | 0.05 | 1.96 | 100 | ** | ** | 42 |
| Si nanowire MZI [14] | Si | Experiment | Electrical | 1.3 | 0.0011 | 0.0018 | 0.61 | 3000 | 80 | 450 | 30 |
| Si nanowire [15] | ITO | 3D FDTD | Electrical | 1.55 | 5.8 | 2 | 2.9 | 1 | 110 | ** | ** |
| Si nanowire [16] | $VO_2$ | Experiment | Electrical | 1.5 | 2.8 | ** | ** | 2.5 | ** | ** | 0.2 |
| Si nanowire micro-ring [17] | $VO_2$ | Experiment | Photothermal | 1.55 | 10 [dB] | ** | ** | 3 (diameter) | 1 | 190 | <10 |
| Si nanowire [18] | Graphene | Experiment | Electrical | 1.55 | 0.104 | 0.076 | 1.37 | 50 | 80 | 350 | 5.9 |
| Cu hybrid plasmonic waveguide [19] | Si | Experiment | Electrical | 1.55 | 1.9 | 1 | 1.9 | 3 | ** | ** | <1 |
| Au plasmonic slot waveguide [20] | ITO | Experiment | Electrical | 1.55 | 2.71 | 0.45 | 6.02 | 10.28 | 50 | 4 | <100 |
| Al hybrid plasmonic waveguide [21] | ITO | 2D mode solving | Electrical | 1.55 | 4.83 | 0.03 | 161 | 0.622 | 270 | 14.8 | 11700 |
| Ag hybrid plasmonic waveguide [4] | $VO_2$ | Experiment | Electrical | 1.55 | 2.3 | 0.86 | 2.67 | 7 | 100 | ** | $400 \times 10^{-9}$ |
| Au hybrid plasmonic near-field coupling [22] | $VO_2$ | 3D FDTD | Electrical | 1.55 | 8.9 | 12.5 | 0.71 | 0.56 | 100 | ** | ~1 |
| **Our design** | $VO_2$ | 3D FDTD | Electrical / Optical | **1.55** | **3.8** | **1.4** | **2.7** | **2** | **>500** | **250** | **~1 (electrical) >100 (optical)** |

Furthermore, the incorporation of phase change materials other than $VO_2$ into the HPSPP waveguide is investigated, and the modulator performance achievable is compared to the $VO_2$ based device.

*A. Engineering E-field Components with HPSPP waveguide*

By switching $VO_2$ from dielectric to metallic, the loss of the modulator goes from low to high, effectively switching from ON- to OFF-state. However, it is useful to gain a deeper understanding of the different contributions to loss: change in permittivity of $VO_2$ and the change in **E**-field intensities within the lossy layers ($VO_2$ and metal). The effective loss of a waveguide $\gamma$ can be expressed in terms of the complex refractive indices of its constituent layers and the distributions of the waveguide mode field components within the waveguide layers, as given by Eq. (1) for the 1D mode [37], which can be extended to the 2D mode in a straightforward manner.

$$\gamma = 2\omega\varepsilon_0 \frac{\int n_r n_i |E|^2 dl}{Re[\int E_y H_x dl]} \quad (1)$$

In Eq. (1), $|E|^2 = E_x^2 + E_y^2 + E_z^2$ is the total electric field intensity, $dl$ is the differential distance along the waveguide cross-section, $\omega$ is the angular frequency of light, and $n_r$ and $n_i$ are the real and imaginary parts of the refractive index, respectively. Note that the direction of propagation along the length of the waveguide is in the $z$-direction. From Eq. (1), it can be seen that the only contribution to waveguide loss is from the layers with non-zero imaginary part of the relative permittivity $\varepsilon_i = 2n_r n_i$. Therefore, it is sufficient to investigate contributions to loss by the intensities of the different electric field components within only each of the lossy layers.

Here we investigate the HPSPP/$VO_2$ modulator using Eq. (1) based on the distribution of EM field components obtained from 1D Lumerical MODE simulations when the top Si layer is varied in thickness, and with the thicknesses of the other material layers kept the same as described previously, namely: 220 nm bottom Si, 20 nm $SiO_2$, 10 nm Ag, and 50 nm $VO_2$ on top. From Fig. 10a, it can be seen that when $VO_2$ is in its dielectric phase, the loss due to the $VO_2$ layer is dominant for low thicknesses of the top Si (high index) layer; however, the loss due to the metal layer increases as the thickness of the high index layer increases, and eventually becomes more dominant compared to the loss due to the $VO_2$ layer for top Si thicknesses >235 nm. When the $VO_2$ layer becomes metallic, its contribution to loss becomes dominant for all top Si thicknesses, and the loss due to the metal layer becomes negligible; this difference in loss contribution between the VO$_2$ and metal layers becomes more pronounced as the thickness of top Si decreases (Fig. 10b). The modulation is thus mainly attributed to the change in loss caused by the VO$_2$ layer when it is switched from dielectric to metallic, with the metal layer playing a very minor role, as shown in Fig. 10c.

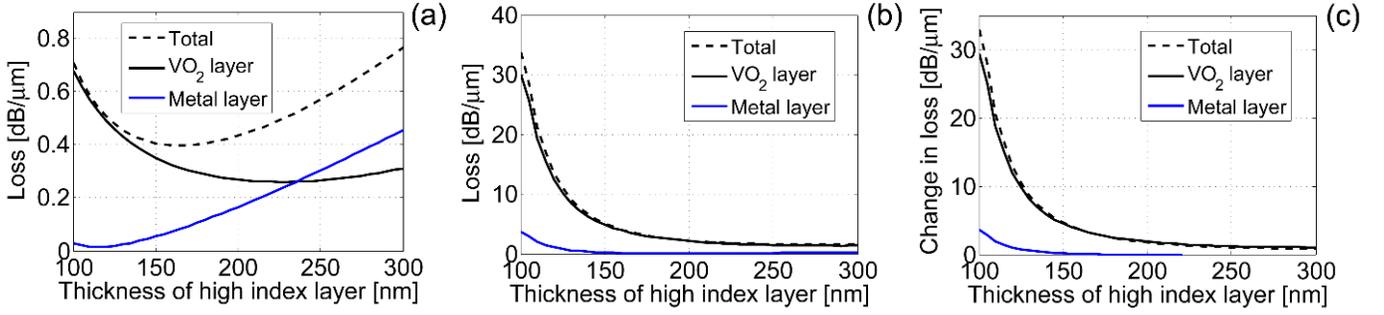

Fig. 10. Loss of the HPSPP/VO$_2$ modulator contributed by each of the VO$_2$ and metal layers, as well as the total, as a function of top Si (high index) layer thickness at λ = 1.55 μm when VO$_2$ is in its (a) dielectric and (b) metallic phase. (c) Change in loss of the HPSPP/VO$_2$ modulator when VO$_2$ is switched from its dielectric to metallic phase, contributed by each of the VO$_2$ and metal layers, as well as the total, as a function of top Si (high index) layer thickness at λ = 1.55 μm (1D mode solving).

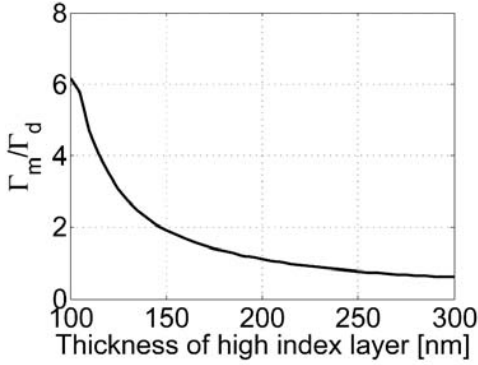

Fig. 11. The ratio of contribution to loss to the HPSPP/VO$_2$ modulator by **E**-field intensity within the VO$_2$ layer between when VO$_2$ is metallic and when it is dielectric ($\Gamma_m/\Gamma_d$), as a function of top Si (high index) layer thickness at λ = 1.55 μm (1D mode solving).

In order to assess the contributions of the change in permittivity of VO$_2$ and the change in **E**-field intensity within the VO$_2$ layer to the modulation process, we introduce the quantity $\gamma_m/\gamma_d$, where $\gamma_m$ and $\gamma_d$ are the losses of the modulator due only to the VO$_2$ layer when it is in its metallic and dielectric phase, respectively. From Eq. (1), it can be deduced that $\gamma_m/\gamma_d$ becomes

$$\frac{\gamma_m}{\gamma_d} = \frac{\varepsilon_{m,i}}{\varepsilon_{d,i}} \times \frac{\left(\frac{\int_{VO_2} |E_m|^2 dl}{Re[\int E_{m,y} H_{m,x} dl]}\right)}{\left(\frac{\int_{VO_2} |E_d|^2 dl}{Re[\int E_{d,y} H_{d,x} dl]}\right)} = \frac{\varepsilon_{m,i}}{\varepsilon_{d,i}} \times \frac{\Gamma_m}{\Gamma_d} \qquad (2)$$

where $\varepsilon_{m,i}$ and $\varepsilon_{d,i}$ are the imaginary part of the relative permittivities of VO$_2$ in its metallic and dielectric phases, respectively; $E_m$ and $E_d$ are the **E**-field distributions when the VO$_2$ layer is in its metallic and dielectric phases, respectively;

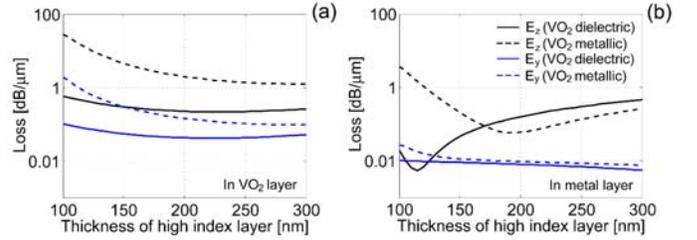

Fig. 12. The contributions to loss of the HPSPP/VO$_2$ modulator by $E_z$ and $E_y$ components of field intensity within the (a) VO$_2$ layer and the (b) metal layer, for when VO$_2$ is in its dielectric and metallic phases, as a function of top Si (high index) layer thickness at λ = 1.55 μm (1D mode solving).

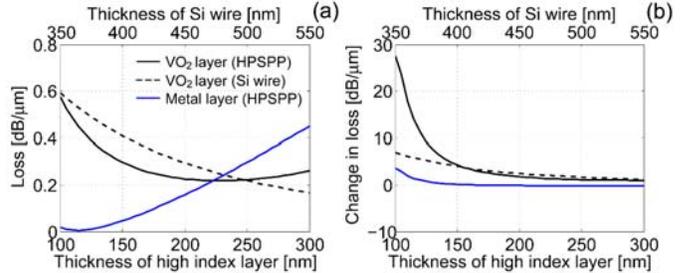

Fig. 13. (a) The contributions to loss of the HPSPP/VO$_2$ and Si wire/VO$_2$ modulators by the $E_z$ field intensities within the VO$_2$ and metal layers when VO$_2$ is in its dielectric phase; and (b) the contributions to change in loss of the HPSPP/VO$_2$ and Si wire/VO$_2$ modulators by the change in $E_z$ field intensities within the VO$_2$ and metal layers when VO$_2$ is switched from its dielectric to metallic phase, as a function of top Si (high index) layer or Si wire thickness at λ = 1.55 μm (1D mode solving).

and $H_m$ and $H_d$ are the **H**-field distributions when the VO$_2$ layer is in its metallic and dielectric phases, respectively. The integration in the numerator of each of $\Gamma_m$ and $\Gamma_d$ is over the VO$_2$ layer only, and the integration in the denominator is over the entire waveguide cross-section. From Eq. (2), the contribution to change in waveguide loss by permittivity change in VO$_2$ is represented by the quantity $\varepsilon_{m,i}/\varepsilon_{d,i}$, and the contribution by the change in **E**-field intensity within the VO$_2$ layer is specified by $\Gamma_m/\Gamma_d$, which is shown in Fig. 11. The ratio $\varepsilon_{m,i}/\varepsilon_{d,i}$ can be obtained from the complex refractive indices of VO$_2$ in its different material phases, and it is ~7.17 at the operating wavelength λ = 1.55 μm. From Fig. 10, it can be observed that at the top Si layer thickness of 150 nm, $\Gamma_m/\Gamma_d$ ≈ 1.92, which means the modulation is predominantly due to the change in complex permittivity in comparison to the change in **E**-field intensity within the VO$_2$ layer. The permittivity of VO$_2$ in each of its phases and thus the ratio $\varepsilon_{m,i}/\varepsilon_{d,i}$ is constant, but the change in **E**-field intensity within the VO$_2$ layer with modulation can be tuned by modifying the waveguide structure. As the top Si layer thickness of the HPSPP/VO$_2$ modulator is decreased, it is seen from Fig. 11 that the contribution to waveguide loss by shifting of **E**-field into the VO$_2$ layer increases, and it becomes comparable to the contribution due to permittivity change for low top Si layer thicknesses close to 100 nm.

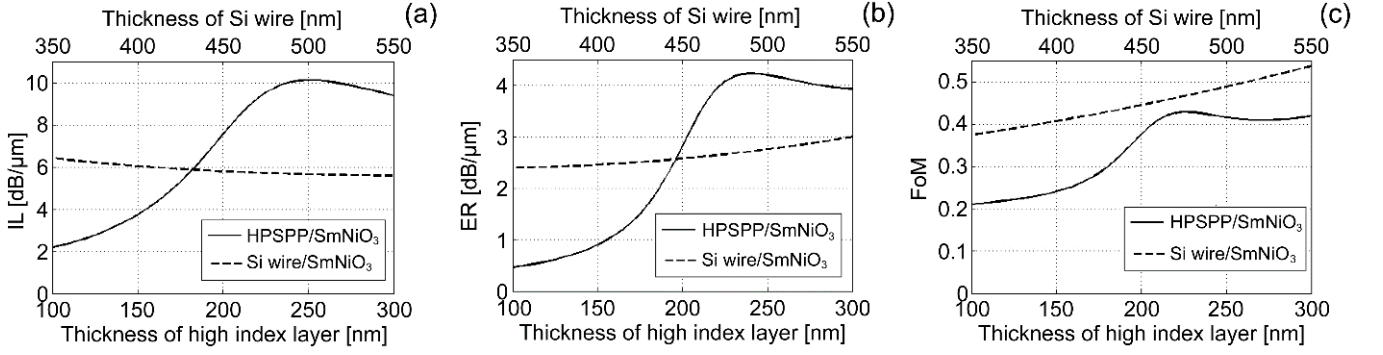

Fig. 14. (a) Insertion loss, (b), extinction ratio, and (c) Figure-of-Merit of HPSPP/SmNiO$_3$ and Si wire/SmNiO$_3$ waveguides as a function of top Si (high index) layer or Si wire thickness (1D mode solving).

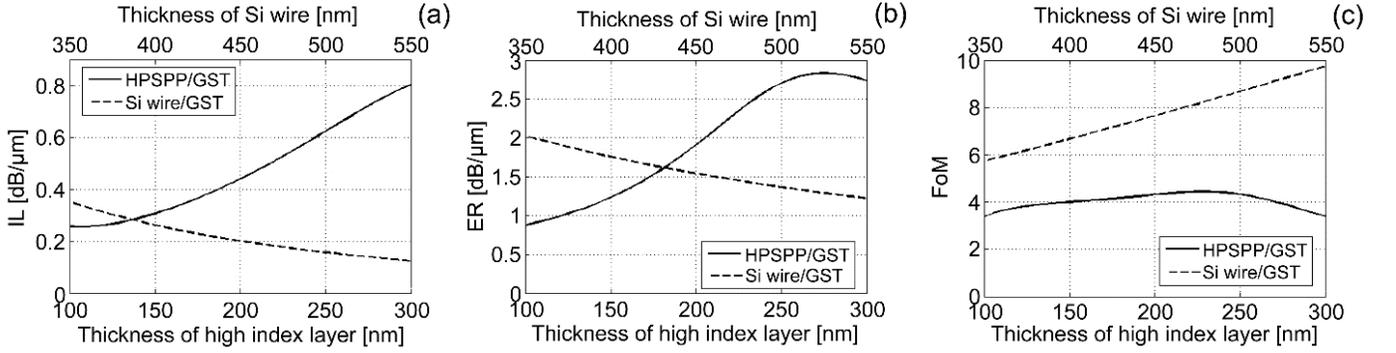

Fig. 15. (a) Insertion loss, (b) extinction ratio, and (c) Figure-of-Merit of HPSPP/GST and Si wire/GST waveguides as a function of top Si (high index) layer or Si wire thickness (1D mode solving).

The **E**-field intensity $|E|^2$ can be separated into its components in the $x$, $y$, and $z$-directions of polarization. As such, the contributions to waveguide loss by each of these **E**-field components within the lossy layers can be investigated. For the HPSPP/VO$_2$ modulator, the $E_x$ field component is approximately 15-orders of magnitude lower than the other two components and thus it is negligible, so here we only present the results for the $E_y$ and $E_z$ field components. From Fig. 12a, it is observed that when VO$_2$ is dielectric, the contribution to waveguide loss by the $E_z$ field within the VO$_2$ layer is close to an order of magnitude higher than by the $E_y$ field component in the same layer, and this is observed over the entire range of top Si layer thicknesses studied. When VO$_2$ is switched to its metallic phase, the contribution by the $E_z$ field in VO$_2$ is over an order of magnitude higher compared to the $E_y$ field; with top Si thickness of 150 nm, the $E_z$ field in the VO$_2$ layer contributes ~4.56 dB/µm to the waveguide loss, while $E_y$ field in the VO$_2$ contributes only ~0.3 dB/µm. It can be seen by comparing Figs. 12a and 12b that, in general, the contribution to loss by the metal layer is much lower than by the VO$_2$ layer, both for the contributions by $E_z$ and $E_y$ components, which is in agreement with the results shown in Fig. 10. An interesting point is that for top Si thicknesses <170 nm, switching VO$_2$ from its dielectric to metallic phase increases the $E_z$ intensity within the metal layer, but conversely for top Si thicknesses >170 nm, the same switching action decreases $E_z$ intensity within the metal layer.

The better performance of the HPSPP/VO$_2$ compared to Si wire/VO$_2$ modulator in terms of ER and FoM can be explained by investigating the loss contributions of the $E_z$ field intensity within the VO$_2$ layer when it is dielectric, and the change in its contribution to loss when VO$_2$ is switched from dielectric to metallic, as shown in Fig. 13. The contribution to waveguide loss by the $E_y$ field is relatively insignificant as described previously. From Fig. 13b, it is observed that the increase in loss due to shifting of the $E_z$ field intensity to the VO$_2$ layer when its phase transitions from dielectric to metallic can be much higher for the case of the HPSPP in comparison to the Si wire, for low thicknesses of the top Si layer (i.e., <120 nm). When the top Si layer thickness is 100 nm, this change in loss by using HPSPP is ~4 times higher than by employing Si wire. The additional metal layer present in the HPSPP but not in the Si wire further contributes to the increase in loss for low top Si layer thicknesses (Fig. 13b). As such, the ER attainable by proper design of the HPSPP/VO$_2$ modulator can significantly surpass that of the Si wire/VO$_2$ counterpart. The remarkable property of the HPSPP/VO$_2$ modulator is that while the ER can be much higher than using Si wire/VO$_2$, the IL can simultaneously be lower, and hence it is beneficial to the FoM. The reason for this is deduced from Fig. 13a, in which the contribution to loss by the $E_z$ field within the VO$_2$ layer when it is dielectric is shown to be lower by utilizing the HPSPP compared to using Si wire for top Si layer thicknesses of the HPSPP of <245 nm. It can also be observed that the $E_z$ field within the metal layer contributes significantly to the loss as well, and thus in order for the overall IL of the HPSPP/VO$_2$ to be lower than that of the Si wire/VO$_2$ modulator, the top Si layer thickness must be approximately <150 nm.

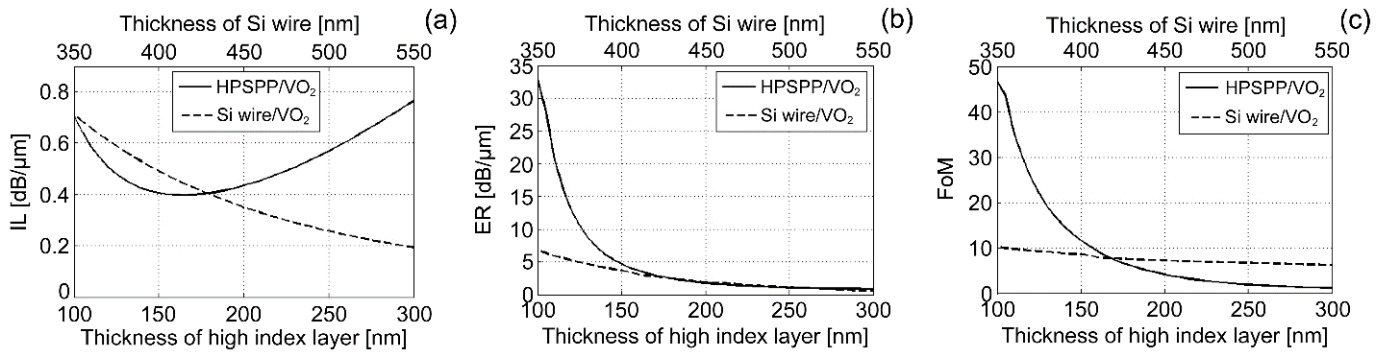

Fig. 16. (a) Insertion loss, (b) extinction ratio, and (c) Figure-of-Merit of HPSPP/VO$_2$ and Si wire/VO$_2$ waveguides as a function of top Si (high index) layer or Si wire thickness (1D mode solving).

*B. HPSPP Modulators Utilizing Other Phase Change Materials*

In this subsection, two other PCMs are used as examples to investigate whether the HPSPP waveguide structure can offer enhanced modulator performance compared to when a TIR waveguide such as Si wire is utilized. This will be investigated through 1D Lumerical MODE simulations [28], and the results are compared to the case when VO$_2$ is used as the switching material. Note that 1D mode solving is sufficient to capture the mode properties of the modulator, especially since we are utilizing the respective predominantly TM modes of the HPSPP and Si wire waveguides, and thus can effectively help make performance comparisons between different devices.

Samarium nickelate (SmNiO$_3$) is a material that sits between strong and weak inter-electron interaction regimes. Like VO$_2$, it is also a correlated electron material that exhibits a sharp reversible insulator-to-metal transition (MIT), and the MIT consists of both electronic and structural changes. The differences of SmNiO$_3$ compared to VO$_2$ are that its phase transition temperature is much higher at 390$^O$C, and there is no hysteresis in the conductivity between switching from insulator to metal and vice versa. The MIT can also be controlled by chemical substitutions, strain, heterostructures, gating, and light pulses [38]. By switching SmNiO$_3$ from its dielectric phase at T = 100 K to metallic phase at T = 390 K, the complex refractive index changes from $n_d$ = 10.09 + 9.90$i$ to $n_m$ = 7.97 + 7.91$i$ at the operating wavelength of $\lambda$ = 1.55 μm. Like the study of the HPSPP/VO$_2$ modulator, SmNiO$_3$ is placed on top of the HPSPP structure, with the modulator ON- and OFF-states corresponding to when SmNiO$_3$ is in its dielectric and metallic phases, respectively. The deposition of SmNiO$_3$ on different substrates including Si, SiO$_2$, and Si$_3$N$_4$ has been demonstrated in previous studies [39-41]. 1D mode solving is performed for varying the top Si thickness while the SmNiO$_3$ layer thickness is kept at 50 nm. The results show that regardless of the top Si thickness which is tuned from 100 nm to 300 nm, the FoM achieved by utilizing the HPSPP waveguide structure is much lower than by simply placing SmNiO$_3$ on top of a Si wire waveguide (Fig. 14c). With the HPSPP/SmNiO$_3$ modulator, the ER can surpass that of the Si wire/SmNiO$_3$ modulator when the top Si thickness is increased past 195 nm, but the IL also steadily increases to be much higher than the Si wire implementation (Fig. 14b and 14a). This shows that the HPSPP waveguide offers no benefit over the use of a simple Si wire waveguide when SmNiO$_3$ is used as the switching material. From Fig. 14, it can be seen that SmNiO$_3$ is inherently not very useful for modulation, due to its ultra-high absorption; for the simple case of a 50 nm SmNiO$_3$ layer placed on top of the Si wire waveguide of 550 nm thickness, the performance is quite poor with FoM = 0.54 and IL = 5.59 dB/μm.

A different class of phase change materials is the chalcogenide Ge$_2$Sb$_2$Te$_5$ commonly referred to as GST, which is employed in optical memory storage devices such as Blu-ray discs. The material transitions from an amorphous to a crystalline state in a fast ~50 ns timescale, and the amorphous phase has a long lifetime of ~200 years [7]. By switching GST from its amorphous to crystalline state, its refractive index changes from $n_a$ = 4.244 + 0.118$i$ to $n_c$ = 7.148 + 1.049$i$ at $\lambda$ = 1.55 μm [7]. It has also been shown in previous work that GST can be deposited on various substrates such as Si [8, 42]. Performing the same analysis as before via 1D mode solving on the HPSPP/GST waveguide, with 50 nm top GST layer and varying the top Si layer thickness, it is found that the modulator performance is not improved compared to the case of placing GST on top of Si wire waveguide. Here, the modulator ON- and OFF-states correspond to when GST is in its amorphous and crystalline phases, respectively. It is clear from Fig. 15c that the FoM is much lower for all top Si thicknesses ranging from 100 nm to 300 nm when HPSPP is utilized. While the ER is increased to be higher than the Si wire implementation with increase of top Si thickness above 180 nm (Fig. 15b), the IL is also much higher than when Si wire is used (Fig. 15a). Although GST has lower absorption in its amorphous phase than VO$_2$ in its dielectric phase, the change in loss when GST is switched to its crystalline phase is much lower than when VO$_2$ is transformed to its metallic state. As such, the overall modulator performance as dictated by the FoM is much inferior to when VO$_2$ is utilized. Even though the imaginary part of the relative permittivity for GST in its amorphous phase ($\varepsilon_i \approx 1$) is significantly lower than VO$_2$ in its dielectric phase ($\varepsilon_i \approx 1.5$), it is still too high for implementing a modulator where the GST is used as the main waveguide material. For the 300 nm thick GST waveguide on top of SiO$_2$ substrate analyzed using 1D mode solving, it is found that when GST is in its amorphous phase, the fundamental TE and TM modes exhibit losses (or IL) of 4.31 dB/µm and 4.84 dB/µm, respectively. These modal losses are much too high for any practical modulators (i.e., the minimum IL for the HPSPP/VO$_2$ modulator is only ~0.4 dB/µm by 1D mode solving as shown in Fig. 16a).

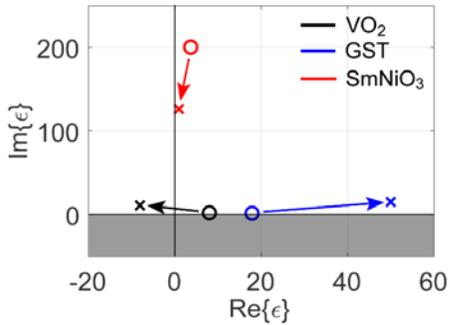

Fig. 17. Plots of the real and imaginary parts of the relative permittivity ε of: VO$_2$ in its dielectric (black circle) and metallic (back cross) phases, GST in its amorphous (blue circle) and crystalline (blue cross) phases, and SmNiO$_3$ in its dielectric (red circle) and metallic (red cross) phases. Only for VO$_2$ does the switching from its material phases transform the material from optically dielectric (Re{ε} > 0) to metallic (Re{ε} < 0). The region of the plot with Im{ε} < 0 is shaded gray because it represents materials with optical gain.

On the contrary, when VO$_2$ is used as the switching material, the HPSPP offers significant modulation performance improvement compared to when the VO$_2$ layer is placed on top of a Si wire waveguide. This has been shown previously by 2D mode solving, but Fig. 16 shows the results from 1D Lumerical MODE simulations for comparison to the cases when other PCMs are used. When the top Si thickness in the HPSPP waveguide is decreased to <150 nm, both the ER and FoM can be dramatically increased in comparison to the Si wire/VO$_2$ waveguide (Fig. 16b and 16c). Simultaneously, the HPSPP/VO$_2$ structure exhibits lower propagation loss when VO$_2$ is in its dielectric state (IL), as shown in Fig. 16a.

By studying Figs. 14, 15 and 16, the best performance (highest FoM) of the HPSPP and Si wire waveguide modulators utilizing VO$_2$, SmNiO$_3$, and GST are extracted and presented in Table 3. It is clearly seen that only by using VO$_2$ as the top layer of the waveguide can the modulator achieve exceptional performance, simultaneously exhibiting high ER and FoM, while still maintaining very low IL. It is interesting to highlight that the maximum ER and FoM achievable with the HPSPP/VO$_2$ modulator are both over 4 times higher than the values attained using the Si wire/VO$_2$ implementation, but both designs exhibit almost identically low IL. However, when either GST or SmNiO$_3$ is utilized, the ER and FoM do not attain very high values. For the case of Si wire/GST, although the FoM attains a reasonably high value of 9.75, the ER is unimpressive at 1.22 dB/µm.

IV. CONCLUSIONS

Phase change materials (PCMs) such as VO$_2$ have the potential to significantly reduce the device footprint of optical modulators, which is an integral component of future optical communications systems that will continue to increase in integration density. The design methodology presented here provides previously unattainable VO$_2$ based modulators that exhibit all of the performance metrics including superior footprint, high ER and FoM, and large optical bandwidth of >500 nm within one device. Furthermore, the projected modulation speed of our HPSPP/VO$_2$ modulator under optical excitation is in the 100's GHz.

Through engineering of **E**-field intensities within the waveguide structure, it is shown that the hybrid plasmonic modulator introduced in this work can significantly improve the performance of VO$_2$ based modulators compared to simply utilizing a TIR waveguide such as a Si wire. This is achieved with the aid of a hybrid plasmonic waveguide that helps maintain a low IL despite of the addition of a lossy metal layer, by drastically reducing the **E**-field (specifically $E_z$) intensity within the VO$_2$ layer, and also maintaining very low **E**-field intensity within the metal layer. Additionally, upon switching VO$_2$ to its metallic state, it is able to significantly increase the field intensity within the VO$_2$ layer compared to the Si wire waveguide implementation, thus achieving a much higher ER and FoM. The VO$_2$ modulator design proposed in this work can be implemented with current technologies in terms of both the multilayer structure and deposition of VO$_2$. We plan to pursue experimental demonstration of our proposed VO$_2$ hybrid plasmonic modulator design in the near

Table 3. Comparison of insertion loss, extinction ratio, and Figure-of-Merit of different HPSPP/PCM and Si wire/PCM modulators investigated in this paper with 50 nm PCM thickness (1D mode solving).

| Modulator device | Top Si [nm] | IL [dB/µm] | ER [dB/µm] | FoM |
|---|---|---|---|---|
| HPSPP/SmNiO$_3$ | 225 | 9.52 | 4.08 | 0.43 |
| Si wire/SmNiO$_3$ | 300 | 5.59 | 3.00 | 0.54 |
| HPSPP/GST | 230 | 0.55 | 2.43 | 4.44 |
| Si wire/GST | 300 | 0.13 | 1.22 | 9.75 |
| HPSPP/VO$_2$ | 100 | 0.71 | 32.98 | 46.60 |
| Si wire/VO$_2$ | 100 | 0.72 | 7.34 | 10.25 |

future.

Using other PCMs, namely SmNiO$_3$ and GST, as the switching material within the hybrid plasmonic modulator (as the top layer) yields device performance similar to that obtainable from simpler implementation using a Si wire waveguide. It is suspected that this is due to the specific way in which the complex permittivity of VO$_2$ changes (i.e., from Re{ε} = 8.11 > 0 to Re{ε} = -8.06 < 0), which is different than with the other investigated PCMs where Re{ε} > 0 (optically dielectric) in both material phases. A summary of the complex permittivity change for the different PCMs when they switch from one material phase to another is depicted in Fig. 17. The nature of the complex permittivity change of the PCM determines the change in the E$_y$ and E$_z$ field intensities within the PCM and metal layers of the HPSPP/PCM waveguide, which in turn dictates the change in modal loss when the PCM is switched.

REFERENCES

[1] D. Dai, J. Bauters, and J. E. Bowers, "Passive technologies for future large-scale photonic integrated circuits on silicon: polarization handling, light non-reciprocity and loss reduction" *Light Sci. Appl.*, vol. 1, e1 doi:10.1038/lsa.2012.1, Mar. 2012.
[2] P. P. Absil, P. Verheyen, P. De Hayne, M. Pantouvanki, G. Lepage, J. De Coster, and J. V. Campenhout, "Silicon photonic integrated circuits: a manufacturing platform for high density, low power optical I/O's," *Opt. Express,* vol. 23, no. 7, pp. 9369–9378, Apr. 2015.
[3] M. Wu, T. J. Seok, and S. Han (2016, Jul.) Large-Scale Silicon Photonic Switches. presented at Advanced Photonics. [Online]. Available: https://www.osapublishing.org/abstract.cfm?uri=SPPCom-2016-JM1A.1
[4] A. Joushaghani, B. A. Kruger, S. Paradis, D. Alain, J. S. Aitchison, and J. K. S. Poon, "Sub-volt broadband hybrid plasmonic-vanadium dioxide switches," *Appl. Phys. Lett.*, vol. 102, no. 6, pp. 061101, Feb. 2013.
[5] Z. Yang and S. Ramanathan, "Breakthroughs in Photonics 2014: Phase Change Materials for Photonics," *IEEE Photon. J.*, vol **7**, no. 3, pp. 0700305, Apr. 2015.
[6] Y. W. Lee, E-S. Kim, B-S. Shin, and S. M. Lee, "High-Performance Optical Gating in Junction Device based on Vanadium Dioxide Thin Film Grown by Sol-Gel Method," *J. Electr. Eng. Technol.*, vol. 7, no. 5, pp. 784-788, May 2012.
[7] N. Yamada and T. Matsunaga, "Structure of laser-crystallized Ge$_2$Sb$_{2+x}$Te$_5$ sputtered thin films for use in optical memory," *J. Appl. Phys.*, vol. 88, no. 12, pp. 7020–7028, Aug. 2000.
[8] C. Rios, P. Hosseini, C. David Wright, H. Bhaskaran, and W. H. P. Pernice, "On-Chip Photonic Memory Elements Employing Phase-Change Materials," *Adv. Mater.*, vol. 26, no. 9, pp. 1372–1377, Mar. 2014.
[9] A. Haché and M. Bourgeois, "Ultrafast all-optical switching in a silicon-based photonic crystal," *Appl. Phys. Lett.*, vol. 77, no. 25, pp. 4089–4091, Dec. 2000.
[10] J. Liu, M. Beals, A. Pomerene, S. Bernardis, R. Sun, J. Cheng, C. Kimerling, and J. Michel, "Waveguide-integrated, ultralow-energy GeSi electro-absorption modulators," *Nat. Photon.*, vol. 2, pp. 433–437, Jul. 2008.
[11] C. Manolatou and M. Lipson, "All-Optical Silicon Modulators Based on Carrier Injection by Two-Photon Absorption," *J. Lightwave Technol.*, vol. 24, no. 3, pp. 1433–1439, Mar. 2006.
[12] B. R. Bennett, R. A. Soref, and J. A. Del Alamo, "Carrier-Induced Change in Refractive Index of InP, GaAs, and InGaAsP," *IEEE J. Quant. Electron.*, vol. 26, no. 1, pp. 113–122, Jan. 1990.
[13] Y. Tang, H-W. Chen, S. Jain, J. D. Peters, U. Westergren, and J. E. Bowers, "50 Gb/s hybrid silicon traveling-wave electroabsorption modulator," *Opt. Express*, vol. 19, no. 7, pp. 5811–5816, Mar. 2011.
[14] M. Streshinsky, R. Ding, Y. Liu, A. Novack, Y. Yang, Y. Ma, X. Tu, E. K. S. Chee, A. E-J. Lim, and P. G-Q. Lo, T. Baehr-Jones, and M. Hochberg, "Low power 50 Gb/s silicon traveling wave Mach-Zehnder modulator near 1300 nm," *Opt. Express*, vol. 21, no. 25, pp. 30350–30357, Dec. 2013.
[15] H. Zhao, Y. Wang, A. Capretti, L. Dal Negro, and J. Klamkin, "Broadband Electroabsorption Modulators Design Based on Epsilon-Near-Zero Indium Tin Oxide," *IEEE J. Sel. Topics Quantum Electron.*, vol. 21, no. 4, pp. 3300207, Aug. 2015.
[16] P. Markov, R. E. Marvel, H. J. Conley, K. J. Miller, R. F. Haglund, Jr., and S. M. Weiss, "Optically Monitored Electrical Switching in VO$_2$," *ACS Photonics*, vol. 2, no. 8, pp. 1175–1182, Jul. 2015.
[17] J. D. Ryckman, V. Diez-Blanco, J. Nag, R. E. Marvel, B. K. Choi, R. F. Haglund, Jr., and S. M. Weiss, "Photothermal optical modulation of ultra-compact hybrid Si-VO$_2$ ring resonators," *Opt. Express*, vol. 20, no. 12, pp. 13215–13225, May. 2012.
[18] Y. Hu, M. Pantouvaki, J. V. Campenhout, S. Brems, I. Asselberghs, C. Huyghebaert, P. Absil, and D. Van Thourhout, "Broadband 10 Gb/s operation of graphene electro-absorption modulator on silicon," *Laser Photonics Rev.*, vol. 10, no. 2, pp. 307–316, Mar. 2016.
[19] S. Zhu, G. Q. Lo, and D. L. Kwong, "Electro-absorption modulation in horizontal metal-insulator-silicon-insulator-metal nanoplasmonic slot waveguides," *Appl. Phys. Lett.*, vol. 99, no. 15, pp. 151114, Oct. 2011.
[20] H. W. Lee, G. Papadakis, S. P. Burgos, K. Chander, A. Kriesch, R. Pala, U. Peschel, and H. A. Atwater, "Nanoscale Conducting Oxide PlasMOStor", *Nano. Lett.*, vol. 14, no. 11, pp. 6463–6468, Oct. 2014.
[21] C. Lin and A. S. Helmy, "Dynamically reconfigurable nanoscale modulators utilizing coupled hybrid plasmonics," *Sci. Rep.*, vol. 5, pp. 12313, Jul. 2015.
[22] P. Markov, K. Appavoo, R. F. Haglund, Jr., and S. M. Weiss, "Hybrid Si-VO$_2$-Au optical modulator based on near-field plasmonic coupling," *Opt. Express*, vol. 23, no. 5, pp. 6878–6887, Mar. 2015.
[23] M. Z. Alam, J. Meier, J. S. Aitchison, and M. Mojahedi, "Propagation characteristics of hybrid modes supported by metal-low-high index waveguides and bends," *Opt. Express*, vol. 18, no. 12, pp. 12971–12979, Jun. 2010.
[24] Y. Su, C. Lin, P. Chang, and A. S. Helmy, "Highly Sensitive Wavelength-scale Amorphous Hybrid Plasmonic Detectors," *Optica*, vol. 4, no. 10, pp. 1259–1262, Oct. 2017.
[25] W. Ma and A. S. Helmy, "Asymmetric long-range hybrid-plasmonic modes in asymmetric nanometer-scale structures," *J. Opt. Soc. Am. B*, vol. 31, no. 7, pp. 1723–1729, Jul. 2014.
[26] H. W. Verleur, A. S. Barker, Jr., and C. N. Berglund, "Optical Properties of VO$_2$ between 0.25 and 5 eV," *Phys. Rev.*, vol. 172, no. 3, pp. 788–798, Aug. 1968.
[27] E. D. Palik, "Handbook of Optical Constants of Solids," vol. 1, Academic Press, Inc., 1985.
[28] Lumerical Solutions, Inc. (www.lumerical.com).
[29] Z. You, C. Xiaonan, K. Changhyun, Y. Zheng, C. Mouli, and S. Ramanathan, "Voltage-triggered ultrafast phase transition in vanadium dioxide switches," *Electron Dev. Lett. IEEE*, vol. 34, no. 2, pp. 220–222, Jan. 2013.
[30] M. Liu, H. Y. Hwang, H. Tao, A. C. Strikwerda, K. Fan, G. R. Keiser, A. J. Sternbach, K. G. West, S. Kittiwatanakul, J. Lu, S. A. Wolf, F. G. Omenetto, X. Zhang, K. A. Nelson, and R. D. Averitt, "Terahertz-field induced insulator-to-metal transition in vanadium dioxide metamaterial," *Nature*, vol. 487, no. 7407, pp. 345–348, Jul. 2012.
[31] A. Pashkin, C. Kübler, H. Ehrke, R. Lopez, A. Halabica, R. F. Haglund, R. Huber, and A. Leitenstorfer, "Ultrafast insulator-metal phase transition in VO$_2$ studied by multiterahertz spectroscopy," *Phys. Rev. B*, vol. 83, no. 19, pp. 195120, May 2011.
[32] A. Liu, L. Liao, D. Rubin, H. Nguyen, B. Ciftcioglu, Y. Chetrit, N. Izhaky, and M. Paniccia, "High-speed optical modulation based on carrier depletion in a silicon waveguide," *Opt. Express*, vol. 15, no. 2, pp. 660–668, Jan. 2007.
[33] M. Rini, A. Cavalleri, R. W. Schoenlein, R. López, L. C. Feldman, R. F. Haglund, Jr., L. A. Boatner, and T. E. Haynes, "Photoinduced phase transition in VO$_2$ nanocrystals: ultrafast control of surface-plasmon resonance," *Opt. Lett.*, vol. 30, no. 5, pp. 558–560, Mar. 2005.
[34] U. Koren, B. I. Miller, T. L. Koch, G. Eisenstein, R. S. Tucker, I. Bar-Joseph, and D. S. Chemia, "Low-loss InGaAs/InP multiple quantum well optical electroabsorption waveguide modulator," *Appl. Phys. Lett.*, vol. 51, no. 15, pp. 1132–1134, Oct. 1987.
[35] D. Marris-Morini, L. Vivien, J. Marc Fédéli, E. Cassan, G. Rasigade, X. Le Roux, P. Lyan, and S. Laval, "Carrier-depletion-based optical modulator integrated in a lateral structure in a SOI waveguide," *Proc. SPIE*, vol. 6996, no. 699615, May 2008.

# Bibliography

[36] E. Rouvalis, C. Metzger, A. Charpentier, T. Ayling, S. Schmid, M. Gruner, D. Hoffmann, M. Hamacher, G. Fiol, and M. Schell, "A Low Insertion Loss and Low $V_\pi$ InP IQ Modulator for Advanced Modulation Formats," in *European Conference on Optical Communication*, paper Tu.4.4.1, Sep. 2014.

[37] A. W. Snyder and J. D. Love, "Optical Waveguide Theory," Chapman and Hall, Ltd., 1983.

[38] S. D. Ha, M. Otaki, R. Jaramillo, A. Podpirka, and S. Ramanathan, "Stable metal-insulator transition in epitaxial $SmNiO_3$ thin films," *J. Solid State Chem.*, vol. 190, pp. 233–237, Jun. 2012.

[39] J. Shi, Y. Zhou and S. Ramanthan, "Colossal resistance switching and band gap modulation in a perovskite nickelate by electron doping," *Nat. Commun.*, vol. 5, no. 4860, Sep. 2014.

[40] Z. Li, Y. Zhou, H. Qi, Q. Pan, Z. Zhang, N. N. Shi, M. Lu, A. Stein, C. Y. Li, S. Ramanathan, and N. Yu, "Correlated Perovskites as a New Platform for Super-Broadband-Tunable Photonics," *Adv. Mater.*, vol. 28, no. 41, pp. 9117 – 9125, Nov. 2016.

[41] R. Jaramillo, F. Schoofs, S. D. Ha, and S. Ramanathan, "High pressure synthesis of $SmNiO_3$ thin films and implications for thermodynamics of the nickelates," *J. Mater. Chem. C*, vol. 1, pp. 2455 – 2462, Feb. 2013.

[42] V. Bragaglia, F. Arciprete, W. Zhang, A. Massimiliano Mio, E. Zallo, K. Perumal, A. Giussani, S. Cecchi, J. Emiel Boschker, H. Riechert, S. Privitera, E. Rimini, R. Mazzarello, and R. Calarco, "Metal-Insulator Transition Driven by Vacancy Ordering in GeSbTe Phase Change Materials," *Sci. Rep.*, vol. 6, no. 23843, Apr. 2016.
**Herman M. K. Wong** received the B.A.Sc. degree from the Division of Engineering Science, with specialization in nanoengineering, in 2008 and the M.A.Sc. degree from the Department of Electrical and Computer Engineering with a focus on integrated photonics and plasmonics in 2011, both from the University of Toronto, ON, Canada, where he is currently working toward the Ph.D. degree in the Photonics Group, Department of Electrical and Computer Engineering. He was a Graduate Research Intern at Intel Labs in Santa Clara, CA, USA in 2014, working on the characterization of silicon photonic devices. He also completed an internship at the Institute of Photonic Sciences in Barcelona, Spain in 2009, conducting research on surface plasmon-based optical tweezers. His research interests include plasmonic devices and systems, metamaterials, silicon photonics, and their applications to optical communications and sensing.

**Amr S. Helmy** received the B.Sc. degree in electronics and telecommunications engineering from Cairo University in 1993, the M.Sc. and Ph.D. degrees from the University of Glasgow with a focus on photonic devices and fabrication technologies, in 1999 and 1995, respectively. He is currently a Professor in the Department of Electrical and Computer Engineering, University of Toronto. Prior to his academic career, he held a position at Agilent Technologies photonic devices, R&D division, U.K. between 2000 and 2004. His research interests include photonic device physics and characterization techniques, with emphasis on nonlinear optics in III–V semiconductors, applied optical spectroscopy in III–V optoelectronic devices and materials, III–V fabrication and monolithic integration techniques.13